\documentclass[usenatbib,twocolumn]{mn2e}
\usepackage{graphicx}
\usepackage{amsfonts}
\usepackage{epsfig,rotating,natbib,pstricks}
\usepackage{color}
\usepackage{amssymb,latexsym}
\usepackage{amsmath,wasysym}
\usepackage{verbatim}
\usepackage[T1]{fontenc}
\usepackage{aecompl}

\def\beq{\begin{equation}}
\def\eeq{\end{equation}}
\def\baq{\begin{eqnarray}}
\def\eaq{\end{eqnarray}}

\def\p3m{P$^3$M}
\def\ap3m{AP$^3$M}

\def\h1{H\/I}
\def\omegah1{\Omega_{\h1}}
\def\ph1{P_{_{\h1}}}
\def\ph1k{P_{_{\h1}}(k)}
\def\dh1k{\Delta^2_{_{\h1}}(k)}

\def\mblack2{{MassiveBlack-II }}
\def\mblack{{MassiveBlack }}

\def\mhi{M_{\text{HI}}}

\def\w50{W_{50}}
\def\s21{S_{21}}

\newcommand{\be}{\begin{equation}}
\newcommand{\e}{\end{equation}}

\title[The $\Omega_{\text{HI}}$ Distribution]{The Distribution of Neutral Hydrogen in 
the Color-Magnitude Plane of Galaxies}

\author[]{\parbox{18cm}{Saili Dutta$^{1}$\thanks{E-mail: sailidutta@niser.ac.in (SD)},
Nishikanta Khandai$^{1}$\thanks{E-mail: nkhandai@niser.ac.in (NK)}}
  \vspace{0.3cm}\\
  $^{1}$ {School of Physical Sciences, 
    National Institute of Science Education and Research, HBNI, Jatni 752050, India}}
\label{firstpage}

\def\LaTeX{L\kern-.36em\raise.3ex\hbox{a}\kern-.15em
    T\kern-.1667em\lower.7ex\hbox{E}\kern-.125emX}

\pagerange{\pageref{firstpage}--\pageref{lastpage}}

\begin{document}

\maketitle

\begin{abstract}
We present the conditional HI (neutral hydrogen) 
Mass Function (HIMF) conditioned on observed 
optical properties, $M_{\text{r}}$ ($r$-band absolute magnitude) and  
$C_{\text{ur}}$ ($u-r$ color), for a sample of 7709 galaxies 
from ALFALFA (40\% data release - $\alpha.40$) which overlaps with a common 
volume in SDSS DR7. Based on the conditional HIMF 
we find that the luminous red, luminous blue and faint blue populations 
dominate the total HIMF at the high-mass end, knee and the low-mass end respectively.
We use the conditional HIMF to derive 
the underlying distribution function of $\Omega_{\text{HI}}$ (HI density parameter), 
$p(\Omega_{\text{HI}})$, 
in the color-magnitude  plane of galaxies.  
The distribution, $p(\Omega_{\text{HI}})$, peaks in the blue cloud at  
$M_{\text{r}}^{\text{max}}=-19.25, C_{\text{ur}}^{\text{max}}=1.44$ but 
is skewed. It has a long tail towards faint blue galaxies and luminous red galaxies.  
We argue that $p(\Omega_{\text{HI}})$ can be used to reveal the underlying relation
between cold gas, stellar mass and the star formation rate (SFR) in an unbiased way; 
that is the derived relation does not suffer from survey or sample selection.  
\end{abstract}



\section{Introduction}
Cold gas represents an important  baryonic component of galaxies 
since it indicates the amount of  gas that is available for future star formation 
of galaxies. Observationally the star formation surface density is strongly 
correlated with the cold gas (neutral hydrogen: sum of atomic, HI, 
and molecular, $\text{H}_2$) surface density in late type disk galaxies --- the
Kennicutt-Schmidt (KS) law \citep{1959ApJ...129..243S,1963ApJ...137..758S,
1998ApJ...498..541K,1989ApJ...344..685K} for star formation. 
Targeted observations have detected HI in late-type (E and S0) 
galaxies  
\citep{2006MNRAS.371..157M,2007A&A...465..787O,2012MNRAS.422.1835S}, 
but their star formation rate is negligible to construct a 
corresponding KS-like law for them. Blind surveys on the other hand
have constrained the HIMF in the local Universe 
\citep{2003AJ....125.2842Z,
2010ApJ...723.1359M,2011AJ....142..170H,2018MNRAS.477....2J}, but the HIMF 
does not reveal how HI is distributed amongst different galaxy populations.

Although the HIMF, and other one-dimensional functions 
(e.g. multiband luminosity functions, stellar mass functions, SFR function, to name a few) 
are  important distributions which any theory of galaxy formation 
should reproduce, they only represent  marginalized distributions of higher dimensional
multivariate distribution functions of galaxies. 
These multivariate functions encode the effects and interplay of complex processes between  
various baryonic components of galaxies. With the advent of ongoing and future 
large surveys which target different bands of the electromagnetic spectrum there is 
a need to go beyond one-dimensional functions. It is common to present bivariate
or multivariate functions, when the observables are from different surveys, as 
conditional functions. 
The bivariate HI mass -- B-band luminosity function 
was estimated from a sample of 61 galaxies in the blind Arecibo HI Strip Survey (AHISS)
\citep{2001MNRAS.327.1249Z}. More recently \cite{2013ApJ...776...74L}
presented the 
HI mass -- stellar mass bivariate function using a parent sample of 480 galaxies 
from the GALEX Arecibo SDSS Survey 
(GASS) Data Release 2 \citep{2010MNRAS.403..683C,2012A&A...544A..65C} 

In this work we present the conditional HIMF conditioned on optical color and magnitude
using a sample of 7709 galaxies from the blind Arecibo Legacy Fast ALFA  (ALFALFA) survey. 
We then use the conditional HIMF to estimate, for the first time, 
the two-dimensional distribution function of $\Omega_{\text{HI}}$ 
in the color-magnitude (CM) plane of galaxies. 
Our paper is organized as follows: we describe our data in section~\ref{sec_data}
followed by a brief description of estimating the HIMF in section~\ref{sec_himf}. We 
present our results in section~\ref{sec_results} and discuss our 
results in section~\ref{sec_discussion}. We assume the following cosmology:
$\left\{ \Omega_{\Lambda}, \Omega_{m}, h \right\} = \left\{0.7,0.3,0.7 \right\}$.

\section{Data}
\label{sec_data}
We give a brief summary of our sample which 
is based on the  $\alpha.40$ data release of ALFALFA 
\citep{2011AJ....142..170H} and is the same as in \cite{2020MNRAS.494.2664D} 
(hereafter, D20). 
We choose an area overlapping with the SDSS DR7 \citep{2009ApJS..182..543A} 
footprint and the 
$\alpha.40$ sample and restrict the redshift range to $cz_{cmb}=15000 km s^{-1}$, to 
avoid RFI.
This common volume is $\sim 2.02 \times 10^6$  Mpc$^3$, 
and subtends an angular area of $\sim 2093$ deg$^2$. 
We also consider only Code 1 objects which have a SNR > 6.5.
Finally we apply the 50\% completeness cut as described in 
\citet{2011AJ....142..170H}, which brings
our final sample to 7857 galaxies. Of these, 7709 galaxies (or 98\%) 
have optical counterparts in SDSS and we loosely refer to the remaining 148 (2\%)
galaxies as \emph {dark} galaxies. In D20 we showed that the dark galaxies 
contribute about $\sim 3\%$ to $\Omega_{\text{HI}}$. Our results should, therefore, 
not be sensitive to this population of dark galaxies.
Of the 7709 galaxies which have optical counterparts in SDSS DR7,
we use their \textit{ugriz} model magnitudes (extinction corrected)
and redshifts to obtain absolute magnitudes  ($M_u,M_g,M_r,M_i,M_z$)
using  \textit{kcorrect} \citep{2007AJ....133..734B}. 
The SDSS galaxy distribution in the CM plane is bimodal.
The dot-dashed curve in figure~\ref{fig_omega_contour}
is the optimal divider to classify these galaxies into red (above curve) and 
blue (below curve) populations \citep{2004ApJ...600..681B}. 
A bimodal distribution is not seen 
in our HI-selected sample (figure~\ref{fig_omega_contour}) because 
ALFALFA primarily samples
the blue cloud, but is nevertheless seen in SDSS for the volume considered here 
(see figure 3 of D20) and we refer to them accordingly as red and blue galaxies.
We restrict our study to the  $\alpha.40$ rather than the recently released
100\% catalog ($\alpha.100$) \citep{2018ApJ...861...49H}. This is because we find 
that at lower declinations, which are now covered by  $\alpha.100$,
many galaxies have luminous foreground stars (as seen in the images) 
and photometric values are not available since  SDSS has masked these regions.
We will consider the $\alpha.100$ sample in the future.

\section{Estimation of HIMF}
\label{sec_himf}
The HIMF, $\phi(M_{\text{HI}})$,  
represents the underlying number density of galaxies in the Universe 
as a function of their HI mass. This is written as 
\beq
\phi(M_{HI}) = \frac{dN}{V \, d\log_{10}M_{HI}} 
\eeq
Here $dN$ is the number of galaxies with masses in the range 
$[\log_{10}M_{\text{HI}},\log_{10}M_{\text{HI}}+d\log_{10}M_{\text{HI}}]$ 
and $V$ is the survey volume of interest.
A single Schechter function has been shown to describe 
the HIMF reasonably well \citep[][D20]{2003AJ....125.2842Z,2010ApJ...723.1359M,
2011AJ....142..170H,2018MNRAS.477....2J}:
\beq
\phi (M_{\text{HI}}) = \ln (10) \; \phi_* \left( \frac{M_{\text{HI}}}{M_*}\right)^{\alpha+1} \exp \left(- \frac{M_{\text{HI}}}{M_*}\right)
\label{eq_schecterfn} 
\eeq
where, $\phi_*$ is the amplitude, $\alpha$ is the slope at the low mass end 
and $M_*$ is the knee of the HIMF, beyond which the 
galaxy counts drop exponentially.
Converting the observed counts of galaxies to the HIMF is non-trivial.
ALFALFA being a blind survey, its sensitivity affects the observed counts.
In the context of ALFALFA the sensitivity limit depends both on the galaxy flux and 
velocity width $W_{50}$. However the data of ALFALFA is large enough 
so that it can be used itself to 
estimate the completeness limit \citep[][]{2011AJ....142..170H}.  
We use the 50\% completeness curve \citep{2011AJ....142..170H} 
as our sensitivity limit.

We use the two-dimensional step wise maximum likelihood 
(2DSWML) method  
\citep{2000MNRAS.312..557L,2003AJ....125.2842Z,2010ApJ...723.1359M,2011AJ....142..170H} 
to estimate the HIMF. The 2DSWML estimator is based on the assumption
that the observed sample of  galaxies is drawn from an underlying distribution function.
In our case it is a bivariate HI mass-velocity width function, $\phi(M_{\text{HI}}, W_{50})$.
The advantage of this method is that it is less susceptible 
to effects of large scale structure (e.g. clustering) 
and the stepwise nature of the method does not assume a functional form but rather 
estimates $\phi_{jk}$.
Here $\phi_{jk} \equiv \phi(M_{\text{HI}}^j,W_{50}^k)$
is the discretised version of bivariate function $\phi(M_{\text{HI}}, W_{50})$ 
in bins of mass, $M_{\text{HI}}^j$ and velocity width $W_{50}^k$. As with any 
maximum likelihood method the normalization is lost and has to be fixed 
separately. We use the method outlined in the appendix of D20 to fix the 
normalization. One can then integrate  $\phi(M_{\text{HI}}^j,W_{50}^k)$
over the velocity width to obtain the HIMF $\phi(M_{\text{HI}}^j)$ 
\citep{2003AJ....125.2842Z, 2010ApJ...723.1359M, 
2011AJ....142..170H, 2018MNRAS.477....2J} or integrate over the mass 
to obtain  $\phi(W_{50}^k)$. 

We estimate errors in the same manner as in D20. The error in mass is related 
to the errors in the observed
flux ($S_{21}$) and the errors in distance ($D$) of each galaxy, since 
$\mhi \propto S_{21} D^2$. Based on the observed values and the estimated errors  
on both flux and distance we generate 300 Gaussian random realizations for each object
in the catalog. These are then used to quote an error for $\phi({M_{\text{HI}}^j)}$.
The second source of errors are Poisson errors
which affects the low and high mass end of the HI catalog, both of which have few objects.
Finally we estimate sample variance by splitting the survey area into  
26 contiguous regions of approximately equal area. We compute the HIMF for each of these
jackknife samples by removing one region at a time. This is then used to compute the 
jackknife uncertainty. One may consider other sources of errors 
\citep[See][]{2018MNRAS.477....2J} but as discussed in D20 these may be  
correlated. For this work we consider the errors outlined above which are consistent with 
\cite{2010ApJ...723.1359M,2011AJ....142..170H}.

\begin{figure*}
  \begin{tabular}{cc}
    \includegraphics[width=3.in]{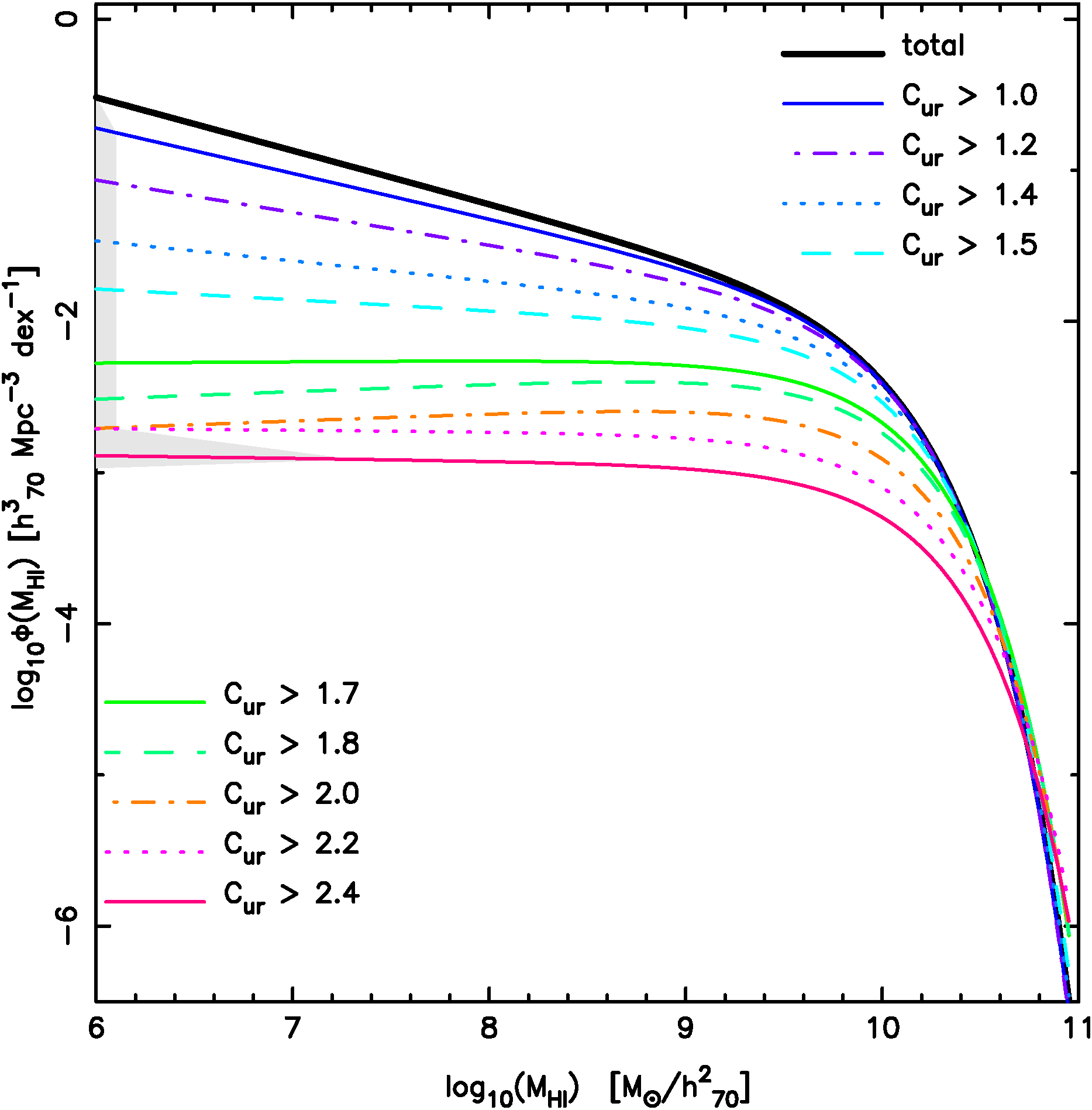}
    \includegraphics[width=3.in]{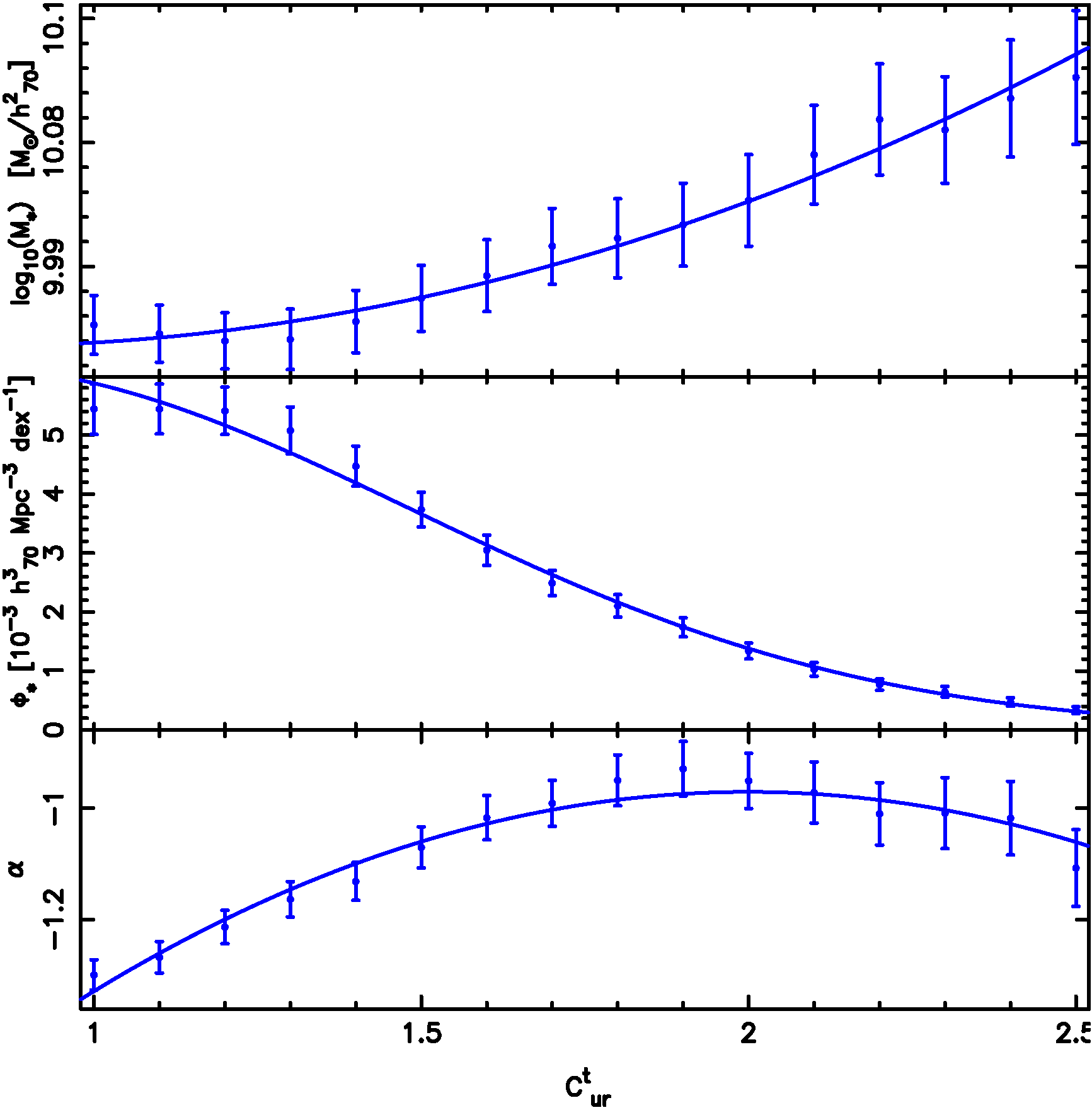}
  \end{tabular}{}
 \caption{\emph{Left}: Conditional HIMF as a function of increasing color thresholds 
(top to bottom). The thick solid line is the HIMF for the full sample.
The shaded gray region does not contain data, the conditional HIMF have however been 
extrapolated into this regime as well. \emph{Right}: The Schechter function parameters of 
the conditional HIMF and their uncertainties as a function of color thresholds. 
The solid lines are fits to the data points with a quadratic function.
The top, middle and bottom panels show the dependence of $M_{*}, \phi_{*}$ and $\alpha$ 
respectively, on the color threshold $C_{\text{ur}}^{\text{t}}$.}
 \label{fig_mf_color}
\end{figure*}

\section{Results}
\label{sec_results}
We now present the results of our paper. Given that 98\% of the HI 
selected galaxies have optical counterparts a natural question would 
be to look at the conditional HIMF, conditioned on an optical property. 
The 2\% of galaxies which are dark, contribute only 3\% to 
$\Omega_{\text{HI}}^{\text{tot}}$ (D20).
In the rest of the paper we will therefore ignore this population 
of dark galaxies since we do not expect them to affect our results quantitatively. 
We emphasize that this is an HI selected sample for which optical properties exist for all 
galaxies. Therefore when computing the HIMF (conditioned on an optical property) 
we need to  consider only the ALFALFA selection function and volume.
In what follows we will compute the HIMF based $M_r$ and $C_{\text{ur}}$ thresholds. 

\subsection{Conditional HIMF}
\label{sec_conditionalmf}
\begin{figure*}
  \begin{tabular}{cc}
 \includegraphics[width=3.in]{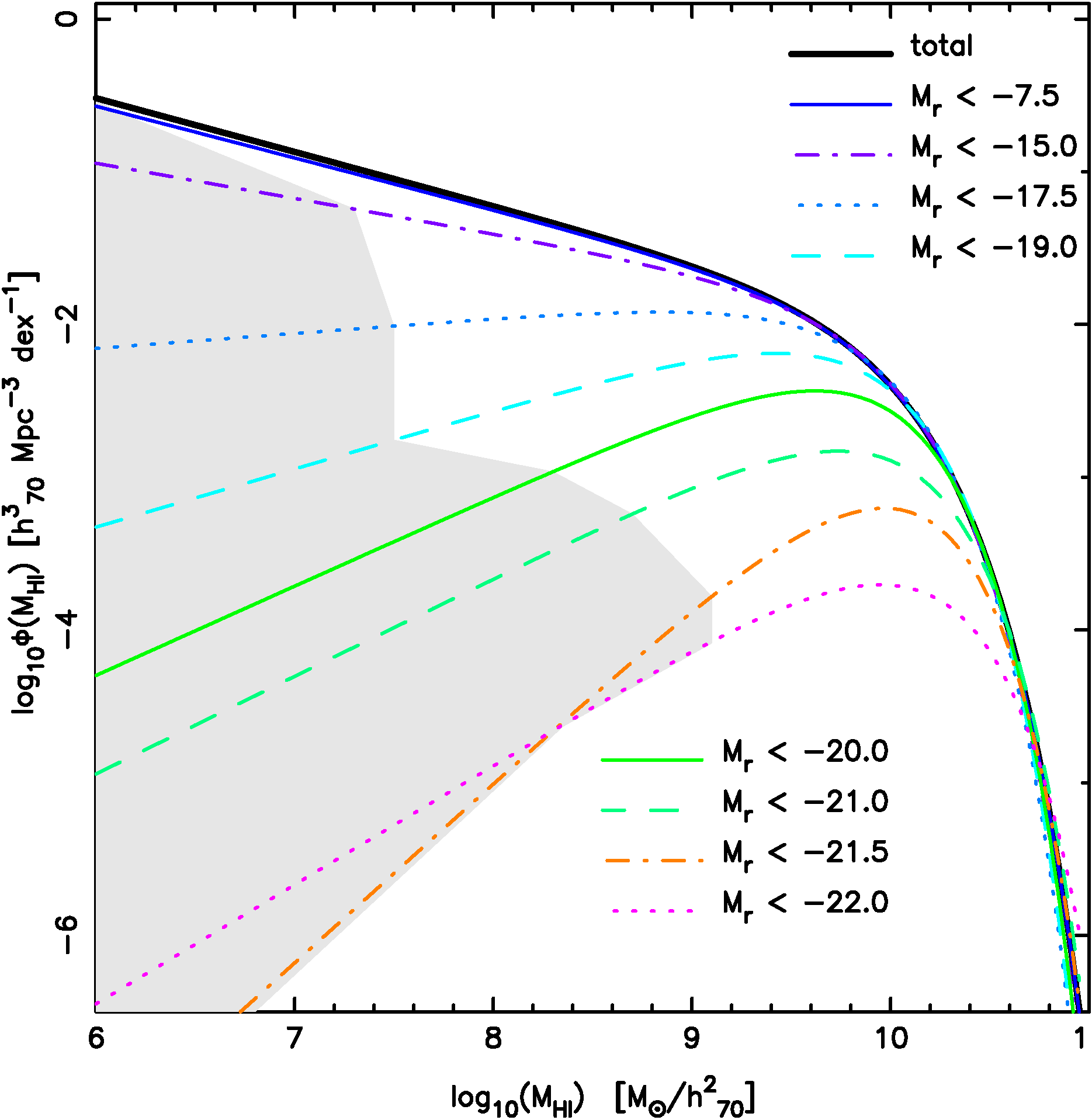}
 \includegraphics[width=3.in]{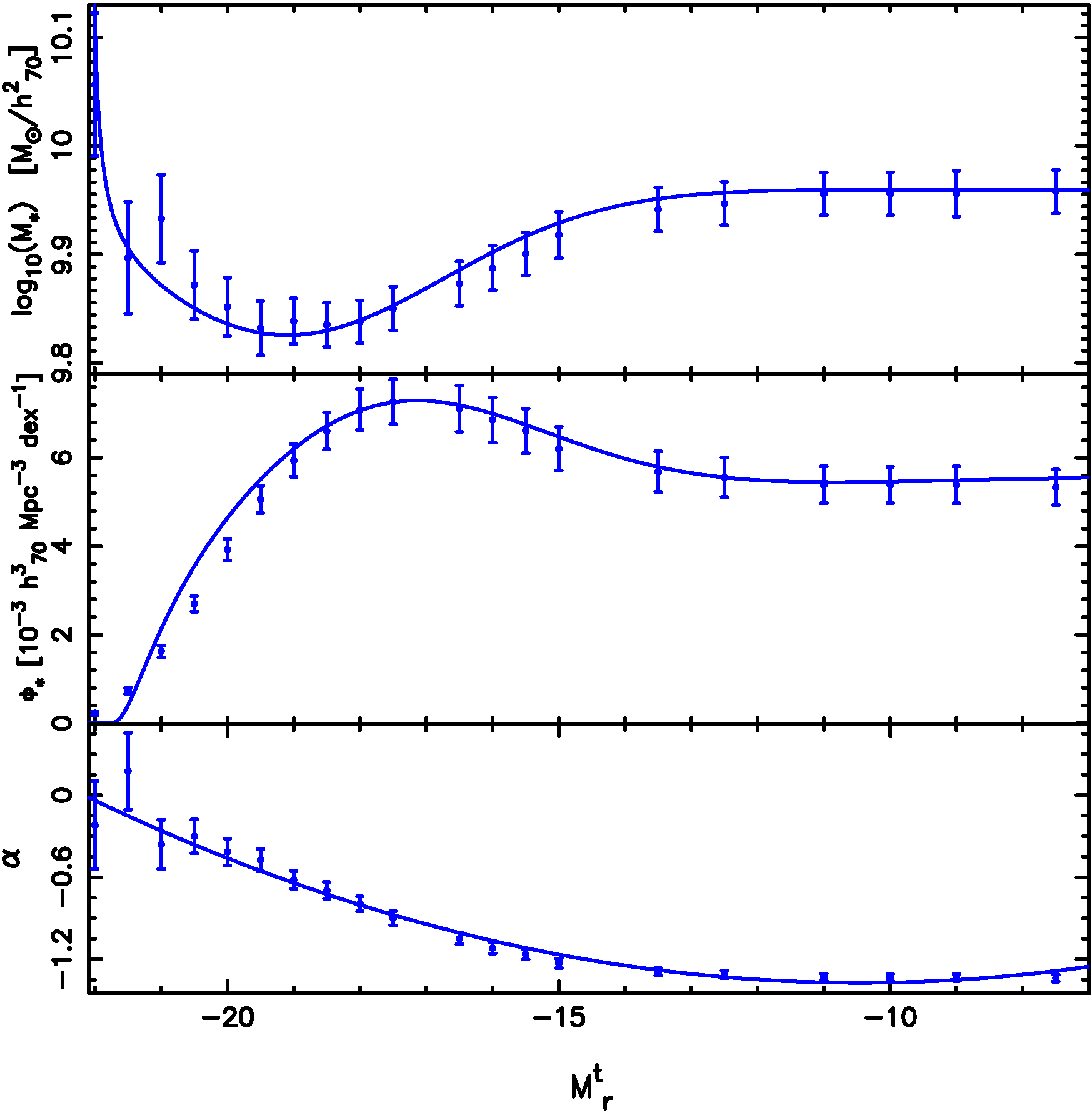}
 \end{tabular}
 \caption{\emph{Left}: Conditional HIMF as a function of decreasing rest frame 
magnitude  thresholds (top to bottom). The thick solid line is the HIMF for the full sample.
The shaded gray region does not contain data, the conditional HIMF have however been 
extrapolated into this regime as well. \emph{Right}: The Schechter function parameters of 
the conditional HIMF and their uncertainties as a function of magnitude thresholds 
$M_{\text{r}}^{t}$. The solid lines are fits to the data points. For $\alpha$ (bottom) 
we fit with a quadratic function. For $M_*$ (top) and $\phi_*$ (middle) we fit 
with a function of the form: 
$y(x) = \left[a + b \exp\left(-\frac{(x+c)^2}{2 d}\right)\right]\frac{f}{(x+e)}$.
\label{fig_mf_mr}}
\end{figure*}

We define the color-conditioned HIMF as    

\beq
\phi(M_{\text{HI}}|C_{\text{ur}}^t) = \phi(M_{HI})|_{C_{\text{ur}} > C_{\text{ur}}^t}
\eeq
This represents the HIMF for galaxies which have a color $C_{ur}$ redder than a threshold 
value $C_{\text{ur}}^t$. Similarly we define the luminosity-conditioned HIMF as
\beq
\phi(M_{\text{HI}}|M_{\text{r}}^t) = \phi(M_{HI})|_{M_{\text{r}} < M_{\text{r}}^t}
\eeq 
which represents the HIMF for galaxies which are more luminous than a threshold 
value $M_{\text{r}}^t$. To compute the conditional HIMF we start 
with the full sample of 7709 galaxies and create a subsample 
based on a threshold color  $C_{\text{ur}}^t$ 
(or magnitude threshold  $M_{\text{r}}^t$). 
We compute the HIMF for this subsample and also estimate its errors as outlined
in section~\ref{sec_himf}. We then fit a Schechter function to obtain a conditional 
HIMF for the particular subsample. 
We repeat this exercise to obtain the conditional HIMF as a function of  
$C_{\text{ur}}^t$ and $M_{\text{r}}^t$. 
Our results are shown in figures~\ref{fig_mf_color} and \ref{fig_mf_mr}.

For the rest of the paper the values of the characteristic 
mass $M_*$ and the amplitude of the Schechter function $\phi_*$ 
(in equation \ref{eq_schecterfn}) will be   
in the units $\left[\log (M_*/M_{\odot}) + 2\log h_{70}\right]$ and  
$\left[10^{-3} h_{70}^{3} Mpc^{-3} dex^{-1}\right]$
respectively. We will also quote $M_{\text{HI}}$ in the same units as $M_*$.

In the left panel of figures~\ref{fig_mf_color} and \ref{fig_mf_mr} 
we show the Schechter function fits to the conditional HIMF. The thick solid line
is the HIMF for the full sample. The shaded gray patch represents 
the region where there is no data. While displaying the Schechter functions we have 
however extrapolated them to this region as well. The right panels represent 
the Schechter function fits and their uncertainties. The lines 
represent a parametric fit to these values. We note that the errorbars on 
the Schechter function parameters, although representative of the sample, 
are correlated, since the sample at each threshold 
(i.e. $C_{\text{ur}}^t$ or $M_{\text{r}}^t$) contains the sample of the previous neighboring
threshold.     

In figure~\ref{fig_mf_color} we look at the color-conditioned HIMF and its dependence 
on the threshold color  $C_{\text{ur}}^t$. 
For  $2.0 \leq C_{\text{ur}}^t \leq 2.4$ the slope at the low mass end is flat, or 
$\alpha \sim -1$ (see bottom right panel of figure~\ref{fig_mf_color}). At this end 
the amplitude, $\phi_*$ is small ($16\times$ smaller) 
compared to the amplitude of the total HIMF, $\phi_* = 5.3\times 10^{-3}$, 
but the characteristic mass $M_* = 10.13$ is about 50\% larger than that of 
the HIMF of the full sample. 
A large value of $C_{\text{ur}}^t$  means
that the subsample contains mostly redder galaxies. 
By decreasing this value  we  add blue galaxies to the sample and 
the conditional HIMF then approaches the total HIMF in the limit 
$C_{\text{ur}}^t \Rightarrow -\infty$. In our sample this is achieved 
when $C_{\text{ur}}^t = 0$. As can be seen in figure~\ref{fig_mf_color} 
there is a near monotonic change in the shape (with the exception of $\alpha$)
of the conditional HIMF with $C_{\text{ur}}^t$. Although $\alpha$ as a function 
of $C_{\text{ur}}^t$ peaks at about $C_{\text{ur}}^t = 1.9$, the variation is still consistent 
with a constant value beyond that. Incidentally the peak in $\alpha$ occurs close to the 
value of the optimal divider of \cite{2004ApJ...600..681B} at $C_{\text{ur}} = 2.3$ 
(see figure~\ref{fig_omega_contour}). 
The red population dominates the HIMF at the large mass end whereas 
decreasing the  $C_{\text{ur}}^t$ we progressively add bluer galaxies to our sample
which start to dominate the knee and then the low mass end for even smaller values of 
$C_{\text{ur}}^t$.

In  figure~\ref{fig_mf_mr} we look at the dependence of  the 
conditional HIMF on $M_{\text{r}}^t$. Unlike the previous case,
the dependence of the conditional HIMF on $M_{\text{r}}^t$ is not monotonic  
(see right panel of figure~\ref{fig_mf_mr}). We see a dip (bump) in $M_*$ ($\phi_*$) 
at $M_{\text{r}}^t = 19$ ($M_{\text{r}}^t \sim 17.5$). 
Coincidentally the distribution of the blue (red) population of galaxies is centered at 
$M_{\text{r}} = 19$ ($M_{\text{r}} = 20$) (See figure 3 of D20). 
As we move from the luminous ($M_{\text{r}}^t \leq 20$ dominated by the red sample), 
to the faint end, the conditional HIMF picks the contribution from the blue cloud at
$M_{\text{r}} = 19$. The bimodality of the underlying optical galaxy sample 
is reflected more strongly in the luminosity-conditioned HIMF than the color-conditioned
HIMF.

\subsection{The Distribution of $\Omega_{HI}$ in the  $C_{\text{ur}}-M_{\text{r}}$ 
  plane}
\label{sec_2domega}
\begin{figure}
 \centering
 \includegraphics[width=\columnwidth]{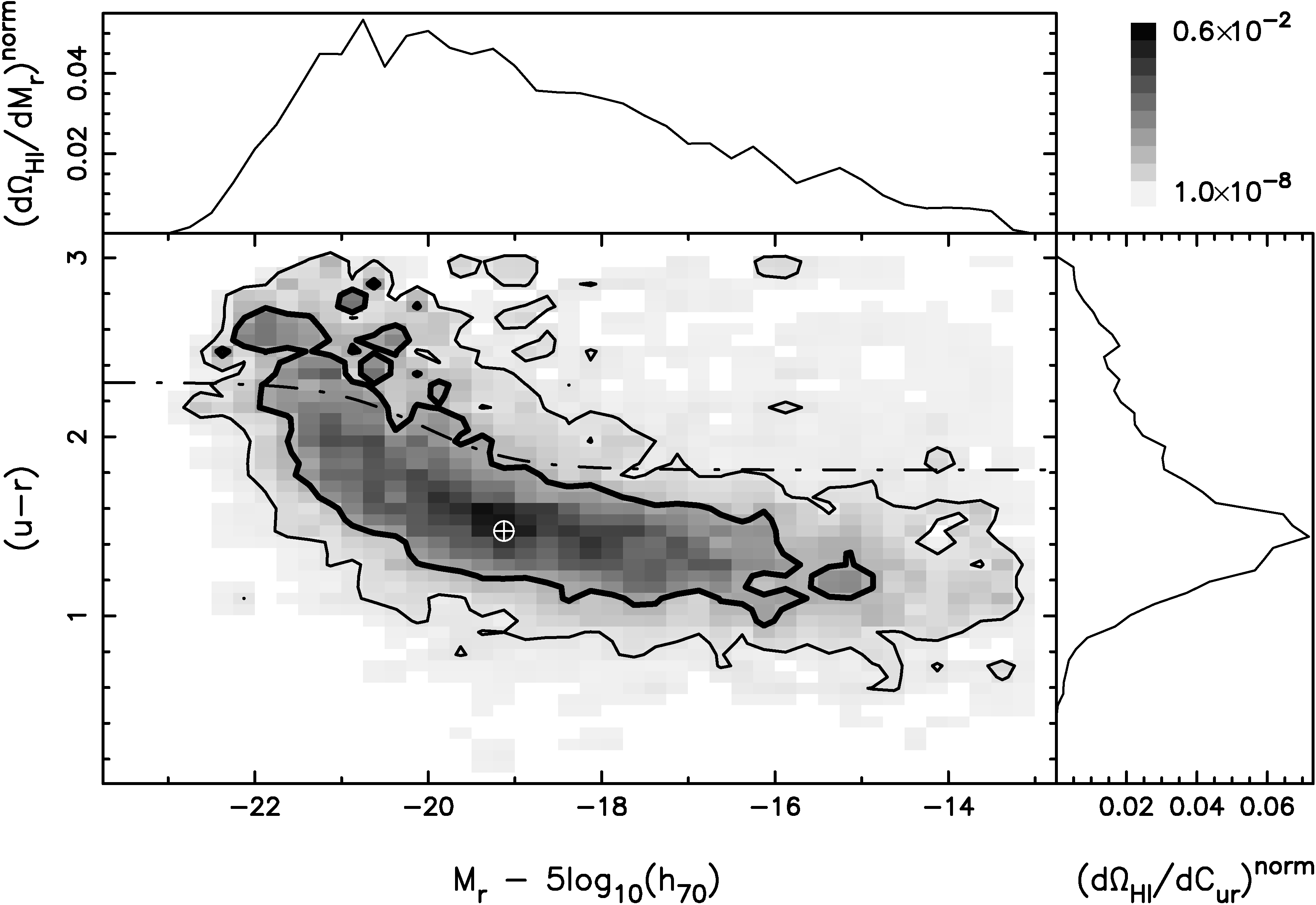}
 \caption{The bottom left panel shows the distribution function  $p(\Omega_{HI})$ 
   (see eq.~\ref{eq_pofomega}) in the CM plane color coded by 
   $(\Omega_{\text{HI}}^{ij})^{\text{norm}}$ (eq.~\ref{eq_omega_ij}). 
   The thick (thin) line   represent the $1\sigma$ ($2\sigma$) widths
   of $p(\Omega_{HI})$.  
   The dash-dot line  separates the optical red (above) and blue (below) populations 
   \citep{2004ApJ...600..681B}. 
   The top left (bottom right) panel is the marginalized distribution of $\Omega_{HI}$ 
   as a function of $M_r$ ($C_{\text{ur}}$ ). The crossed circle represents the peak 
   of the two-dimensional distribution function, $p(\Omega_{HI})$. 
}
 \label{fig_omega_contour}
\end{figure}

We extend our previous definition
to the two-dimensional conditional HIMF:
\beq
\phi(M_{\text{HI}}|C_{\text{ur}}^t,M_{\text{r}}^t) = 
\phi(M_{HI})|_{(C_{\text{ur}} > C_{\text{ur}}^t), (M_{\text{r}} < M_{\text{r}}^t)} 
\eeq
This represents the HIMF of galaxies redder than $C_{\text{ur}}^t$ and more luminous than
$M_{\text{r}}^t$, for which the corresponding HI density parameter is:  
\beq
\Omega_{\text{HI}}(C_{\text{ur}}^t,M_{\text{r}}^t) = \frac{1}{\rho_c}
\int_{0}^{\infty} M_{\text{HI}} 
\phi(M_{\text{HI}}|C_{\text{ur}}^t,M_{\text{r}}^t) 
dM_{\text{HI}} 
\label{eq_omegahi_2d}
\eeq
In our sample 
$\Omega_{\text{HI}}(C_{\text{ur}}^t,M_{\text{r}}^t)=
\Omega_{\text{HI}}^{\text{tot}}=
4.894 \times 10^{-4}$
when $C_{\text{ur}}^t  = 0.0, M_{\text{r}}^t = -6.0$. 
We compute  $2500$ conditional HIMFs and their associated errors
in the CM plane by dividing $C_{\text{ur}}^t \in [3.0, 0.0]$ 
(decreasing color threshold)  
and $M_{\text{r}}^t \in [-23.0, -6.0]$ (increasing magnitude threshold) 
into 50 bins each. From equation~\ref{eq_omegahi_2d} 
we see that the variation of $\Omega_{\text{HI}}(C_{\text{ur}}^t,M_{\text{r}}^t)$
is that of a cumulant in the two-dimensional CM plane. If we define
the normalized conditional HI density parameter as 
$\Omega_{\text{HI}}^{\text{norm}}(C_{\text{ur}}^t,M_{\text{r}}^t)
=\frac{\Omega_{\text{HI}}(C_{\text{ur}}^t,M_{\text{r}}^t)}
{\Omega_{\text{HI}}^{\text{tot}}}$, 
then $\Omega_{\text{HI}}^{\text{norm}}(C_{\text{ur}}^t,M_{\text{r}}^t)$ is bounded 
and varies from 0 (luminous-red, top left corner of figure~\ref{fig_omega_contour})
and 1 (faint-blue, bottom right corner of figure~\ref{fig_omega_contour}).

We define the distribution function of the cosmological HI density parameter  
in the CM plane
\beq
p\left(\Omega_{\text{HI}}(C_{\text{ur}},M_{\text{r}})\right) =  
\left.\frac{\partial^2\Omega_{\text{HI}}^{\text{norm}}(C_{\text{ur}}^t,M_{\text{r}}^t)}
{\partial C_{\text{ur}}^t \partial M_{\text{r}}^t}\right |_
{
[C_{\text{ur}}^t = C_{\text{ur}}, 
M_{\text{r}}^t = M_{\text{r}}] 
\label{eq_pofomega}
}
\eeq
By construction this is a normalized distribution
\beq
\int \int p\left(\Omega_{\text{HI}}(C_{\text{ur}},M_{\text{r}})\right) 
\text{d}C_{\text{ur}} \text{d}M_{\text{r}}  = 1.0
\label{eq_norm_pofomega}
\eeq
The cosmological HI density in a given CM (\emph{ji}) pixel is  
\beq 
(\Omega_{\text{HI}}^{ij})^{\text{norm}} = 
\int_{M_{\text{r}}^i}^{M_{\text{r}}^{i+1}} 
\int_{C_{\text{ur}}^j}^{C_{\text{ur}}^{j+1}}  
p\left(\Omega_{\text{HI}}(C_{\text{ur}},M_{\text{r}})\right) 
\text{d}C_{\text{ur}} \text{d}M_{\text{r}} 
\label{eq_omega_ij}
\eeq    

In figure~\ref{fig_omega_contour} we plot the distribution function, $p(\Omega_{\text{HI}})$,
of the cosmological HI density parameter in the CM plane.
Each pixel is color-coded to the $(\Omega_{\text{HI}}^{ij})^{\text{norm}}$ value. 
The top left (bottom right) panel shows the marginalized distribution 
of $\Omega_{\text{HI}}$ as a function of magnitude (color). The dot-dashed line 
is the optimal divider which classifies these galaxies into red and 
blue populations \citep{2004ApJ...600..681B}. The thick (thin) contour is the 
$1\sigma$ ($2\sigma$) width of the distribution function, $p(\Omega_{\text{HI}})$ (i.e.
the contour is determined from eq.~\ref{eq_norm_pofomega} by setting the RHS to 
0.68 (0.95)).  The crossed-circle is the peak  of $p(\Omega_{\text{HI}})$ in two dimensions,
and does not match the peak of the marginalized distribution because it is skewed.

\section{Discussion}
\label{sec_discussion}
In this paper we have presented the conditional HIMF, 
conditioned on color and/or magnitude. 
Based on the conditional HIMF we obtained the distribution of $\Omega_{\text{HI}}$,
$p(\Omega_{\text{HI}})$, in the CM plane of galaxies. Not surprisingly our results
for  $\phi(M_{HI})|_{M_{\text{r}} < -21}$ and even brighter thresholds is similar to those 
obtained for the conditional HIMF,  $\phi(M_{HI})|_{M_{\text{star}} \geq 10}$, for massive 
galaxies \citep{2013ApJ...776...74L} 
from the GASS survey \citep{2010MNRAS.403..683C,2012A&A...544A..65C};
this is because the stellar mass of galaxies is correlated with its luminosity.

Both the two-dimensional and marginalized distributions 
show that they have long tail towards faint blue galaxies and luminous red galaxies. 
The peak of $p(\Omega_{\text{HI}})$ in the CM plane occurs at 
$C_{\text{ur}}^{\text{max}} = 1.44, M_{\text{r}}^{\text{max}} = -19.25$ 
in the blue cloud, 
which is about 1.36 mag fainter than the characteristic luminosity of blue galaxies 
in SDSS \citep{2004ApJ...600..681B}. 
The width of $p(\Omega_{\text{HI}})$
is also fairly broad in both color and magnitude; 
the average $1\sigma$ ($2\sigma$) widths being $\sigma_C = 0.8, \sigma_M = 3.0$ 
($\sigma_C = 1.1, \sigma_M = 4.8$).  
At the fainter end $M_{\text{r}} > -16$, 
$\sim 10\%$ of $\Omega_{\text{HI}}^{\text{tot}}$ 
is locked in gas rich low surface brightness galaxies. The red population, 
on the other hand, contributes $\sim 18\%$ to the HI budget.

The CM plane can be thought of as a coordinate system in which we can plot 
distributions of other cosmological density parameters (related to galaxies), 
$p(\Omega_{\text{X}})$ where X denotes a property, e.g.  stellar mass 
$M_{\text{star}}$, $SFR$, molecular hydrogen mass $M_{\text{H}_2}$, 
which in turn are computed from $\phi(X|C_{\text{ur}}^t,M_{\text{r}}^t)$. 
We therefore have all the information needed to obtain 
the mean relation between different galaxy properties by discarding the common 
coordinate system. We emphasize that this relation is unbiased and represents 
the underlying relation since the distributions have folded in the survey selection.
The blind nature of the survey is also important since there is no selection bias 
in estimating  $\phi(X)$.
This can be repeated for different galaxies populations (blue or red) and for other bands
as well. 
The methods outlined in this paper are statistical in nature and provide a powerful and 
unbiased way to probe the multivariate distributions of galaxy populations. We will report 
on the mean HI-stellar mass relation in a forthcoming paper. 

\section*{ACKNOWLEDGMENTS}
We would like to thank  R. Srianand, A. Paranjape and J. S. Bagla for useful 
discussions. NK acknowledges the support of the Ramanujan 
Fellowship\footnote{Awarded by the 
Department of Science and Technology, Government of India} 
and  the IUCAA\footnote{Inter University Centre for Astronomy and 
Astrophysics, Pune, India} associateship programme. 
All the analyses were done on the {\tiny{\bf XANADU}} and {\tiny{\bf CHANDRA}} servers 
funded by the Ramanujan Fellowship.

We thank the entire ALFALFA collaboration 
in observing, flagging, and extracting the properties of galaxies that this paper 
makes use of. This work also uses data from SDSS DR7.
Funding for the SDSS and SDSS-II has been provided by the Alfred P. Sloan Foundation, 
the Participating Institutions, the National Science Foundation, 
the U.S. Department of Energy, the National Aeronautics and Space Administration, 
the Japanese Monbukagakusho, the Max Planck Society, and the Higher 
Education Funding Council for England. The SDSS Website is 
http://www.sdss.org/. The SDSS is managed by the Astrophysical Research 
Consortium for the Participating Institutions. 

\section*{Data Availability}
The data used in this work is publicly available.
SDSS DR7 \citep{2009ApJS..182..543A}
data can be accessed from \emph{sciserver.org}
and the $\alpha.40$ \citep{2011AJ....142..170H} 
data from ALFALFA can be accessed from 
\emph{egg.astro.cornell.edu}.

\end{document}